\documentclass[aps,10pt,prl,twocolumn,superscriptaddress]{revtex4-1}

\usepackage[utf8]{inputenc}
\usepackage[T1]{fontenc}
\usepackage[english]{babel}

\usepackage{amsmath}
\usepackage{amssymb}
\usepackage{amsthm}
\usepackage{mathrsfs}

\usepackage{graphicx}
\graphicspath{{Figures/}}

\usepackage{color}
\newcommand{\ericemph}[1]
{
}

\renewcommand{\vec}{\mathbf}

\begin{document}
\title{Universal non-Debye scaling in the density of states of amorphous solids}
\author{Patrick Charbonneau}
\affiliation{Department of Chemistry, Duke University, Durham,
North Carolina 27708, USA}
\affiliation{Department of Physics, Duke University, Durham,
North Carolina 27708, USA}

\author{Eric I. Corwin}
\affiliation{Department of Physics, University of Oregon, Eugene, Oregon 97403, USA}

\author{Giorgio Parisi}
\affiliation{Dipartimento di Fisica,
Sapienza Universit\`a di Roma,
INFN, Sezione di Roma I, IPFC -- CNR,
Piazzale Aldo Moro 2, I-00185 Roma, Italy}

\author{Alexis Poncet}
\email{alexis.poncet@ens.fr}
\affiliation{Department of Physics, University of Oregon, Eugene, Oregon 97403, USA}
\affiliation{D\'epartement de Physique,
\'Ecole Normale Sup\'erieure, 24 Rue Lhomond, 75005 Paris, France}

\author{Francesco Zamponi}
\affiliation{LPT,
\'Ecole Normale Sup\'erieure, UMR 8549 CNRS, 24 Rue Lhomond, 75005 Paris, France}

\begin{abstract}
At the jamming transition, amorphous packings are known to display anomalous vibrational modes with a density of states (DOS) that remains constant at low frequency. The scaling of the DOS at higher densities remains, however, unclear.  One might expect to find simple Debye scaling, but recent results from effective medium theory and the exact solution of mean-field models both predict an anomalous, non-Debye scaling. 
Being mean-field solutions, however, these solutions are only strictly applicable to the limit of infinite spatial dimension, and it is unclear what value they have for finite-dimensional systems. Here, we study packings of soft spheres in dimensions 3 through 7 and find, far from jamming, a universal non-Debye scaling of the DOS that is consistent with the mean-field predictions. We also consider how the soft mode participation ratio converges to the mean-field prediction as dimension increases.
\end{abstract}

\maketitle

\paragraph{Introduction--}
The vibrational spectrum of amorphous solids has long been puzzling materials physicists~\cite{Ph81}. Since their excess low-frequency excitations (compared to crystals) was carefully studied by Raman and neutron scattering~\cite{BND84,BPNDAP86,MS86}, the so-called boson peak anomaly has indeed been given a variety of explanations, ranging from specific features of interatomic forces to a broadened van Hove singularity~\cite{Zorn11}. From the viewpoint of amorphous solids as the paragon of disorder, various models have also been advanced~\cite{GMPV03,SRS07,parisi14,manning15}.
Here, we consider a proposal that recently emerged from the study of the simplest
model of amorphous solids: a disordered assembly of soft, purely repulsive spheres at zero temperature under a confining pressure $P$. This jammed solid becomes mechanically unstable at a sharply defined jamming transition, upon reaching $P=0$~\cite{OSLN03}. The study of this transition by statistical and soft-matter physics~\cite{dynhetjam10} has revealed that the geometric~\cite{morse14}, rheological~\cite{ikeda13}, vibrational~\cite{WSNW05, silbert05} and elastic properties~\cite{ohern03} of solids at jamming markedly differ from those of crystals. In particular, precisely at the jamming transition the density of vibrational states $D(\omega)$, with frequency $\omega$,
becomes flat for $\omega\to 0$, which leads to a diverging boson peak~\cite{silbert05,XWLN07}.
It is thus natural to wonder what is the low-frequency behavior of $D(\omega)$ in the vicinity of this transition, and whether it could provide a universal explanation of the boson peak.

A crucial concept associated with jamming is that of {\it marginal stability}. At the jamming transition the system is on the verge of mechanical instability,
which naturally gives rise to low-energy modes~\cite{ohern03,silbert05}. Surprisingly, it was recently shown that amorphous solids remain marginally stable even at finite
pressures, and, by means of effective medium theory, that this marginality leads to a modified Debye behavior with
$D(\omega) \sim \omega^2$--as in crystals, but with a constant prefactor much larger than expected from standard
elasticity~\cite{degiuli14}. This result offers a promising account for the boson peak. Interestingly, the same scaling behavior was also recently uncovered in the perceptron, which is an exactly solvable model in the same universality class as soft spheres close to jamming~\cite{FPUZ15}. This concordance likely results from both effective medium theory
and the perceptron being mean-field descriptions that are expected to exactly capture the behavior of infinite-dimensional systems.

Before considering possible shortcomings of such descriptions, let us first detail their predictions.
Away from jamming, at large length
scales a solid should behave as a continuum medium, hence one expects a Debye scaling of the density of vibrational states (DOS), i.e., $D(\omega)\sim\omega^{d-1}$, at low frequency.
Effective medium theory~\cite{degiuli14} and the exact solution of the perceptron~\cite{FPUZ15} indeed suggest that
\begin{equation}\label{eq:WyartDOS}
D(\omega) \sim
	\begin{cases}
	\omega^{d-1} & \omega \ll \omega_0 \\
	\omega^2/\omega_\ast^2  & \omega_0 \ll \omega \ll \omega_\ast \\
	\text{constant} & \omega \gg \omega_\ast
	\end{cases},
\end{equation}
where $\omega_\ast$ is a characteristic frequency that vanishes at jamming, 
and $\omega_0$ is a threshold frequency that separates 
the Debye from the anomalous $\omega^2$ regime.
Interestingly, for infinite-dimensional, marginally stable systems, 
${\omega_0=0}$ for a finite region around jamming~\cite{FPUZ15}.
Note that in $d=3$ the Debye and anomalous regime both scale as $\omega^2$,
but the latter has a prefactor, $1/\omega_\ast^2$, that diverges at jamming, hence being much larger
than in Debye's model~\cite{degiuli14}.

Validating these predictions is, however, non-trivial.
For obvious physical reason, most studies of amorphous solids have explored the nature of excitations in two- or three-dimensional systems. These results, however, may be strongly influenced by low-dimensional effects that are absent from mean-field descriptions and may partially obfuscate the universality of the phenomenon. For instance, some of the low-frequency excitations are known to be spatially quasilocalized~\cite{widmer-cooper08,XVLN10}, while only purely delocalized modes are found in the perceptron~\cite{FPUZ15}. Quasilocalized modes are especially significant, because they are associated with structural soft spots~\cite{manning11}. The finite-dimensional behavior of $\omega_0$ may also be richer than in infinite-dimensional models. In this letter, we thus bridge the gap between physical systems and mean-field theories by studying the vibrational modes of soft-sphere packings both as a function of density and of spatial dimension. In doing so, we disentangle universal features from low-dimensional effects in the vibrational spectrum, similarly to what has been done for the force network~\cite{CCPZ12,CCPZ15}.
We also remarkably find that the $\omega^2$ regime is present in all dimensions down to the lowest numerically accessible frequencies.

\paragraph{Model description and Hessian--} We generate packings of $N=8192$ frictionless spheres interacting via a one-sided (contact) harmonic potential within a periodic cubic box in $d=3$--$7$. The total system energy is $U = \frac{1}{2} \sum_{i<j}^N \Theta (\sigma - r_{ij})\left(\sigma - r_{ij}\right)^2$, where $\sigma$ is the particle diameter, $r_{ij}$ is distance between particles $i$ and $j$, and $\Theta$ is the Heaviside step function. The relevant control parameter is the packing fraction, $\phi=\rho V_d(\sigma)$, where $\rho$ is the number density and $V_d(\sigma)$ is the $d$-dimensional volume of a ball of diameter $\sigma$. Initializing with Poisson-distributed spheres at very high $\phi$, configurations are obtained by iteratively (i) deflating particles in small steps and (ii) minimizing the system energy, using the numerical scheme described in Refs.~\cite{CCPZ15,bitzek06}. For $\phi>\phi_\mathrm{J}$ we obtain jammed packings with locally minimal $U>0$, while for $\phi<\phi_\mathrm{J}$ configurations are unjammed, and hence not mechanically stable at $T=0$. The limit case $\phi=\phi_\mathrm{J}$ and $U=0$ is the jamming transition for a given initial configuration. We thus define the excess packing fraction $\Delta\phi\equiv\phi-\phi_\mathrm{J}$. (For this system pressure $P\propto\sqrt{U}\propto\Delta\phi$~\cite{OSLN03,CCPZ15}.) 

In order to extract information about the harmonic excitations of the system, we compute the Hessian matrix
\begin{align*}\label{eq:hessian}
H_{ij}^{\alpha\beta} = \frac{\partial^2 U}{\partial r_i^\alpha \partial r_j^\beta}=&\delta_{ij} \sum_{k \in \partial i} \left[ n_{ik}^\alpha n_{ik}^\beta + \frac{\varepsilon_{ik}}{\rho_{ik}}\left( n_{ik}^\alpha n_{ik}^\beta - \delta^{\alpha\beta}\right)\right] \\
&- \delta_{\langle ij\rangle}\left[ n_{ij}^\alpha n_{ij}^\beta + \frac{\varepsilon_{ij}}{\rho_{ij}}\left( n_{ij}^\alpha n_{ij}^\beta - \delta^{\alpha\beta}\right)\right],
\end{align*}
where $\alpha, \beta = 1\dots d$ are vector components, $\varepsilon_{ij} = \sigma - r_{ij}$ is the overlap between two spheres, $\vec{n}_{ij} = (\vec{r}_j - \vec{r}_i)/r_{ij}$ is a unit vector, both $\delta_{ij}$ and $\delta^{\alpha\beta}$ are Kronecker deltas, $\delta_{\langle ij\rangle}$ indicates a contact between a pair of particles, and $\partial i$ denotes the set of neighbors of $i$. The eigenvectors $\{u_i^\alpha\}_k$ and eigenvalues $\lambda_k$ of the Hessian then provide the vibrational modes and their angular frequencies, $\omega_k = \sqrt{\lambda_k}$, respectively.

\begin{figure}
\centering
\includegraphics[width=\columnwidth]{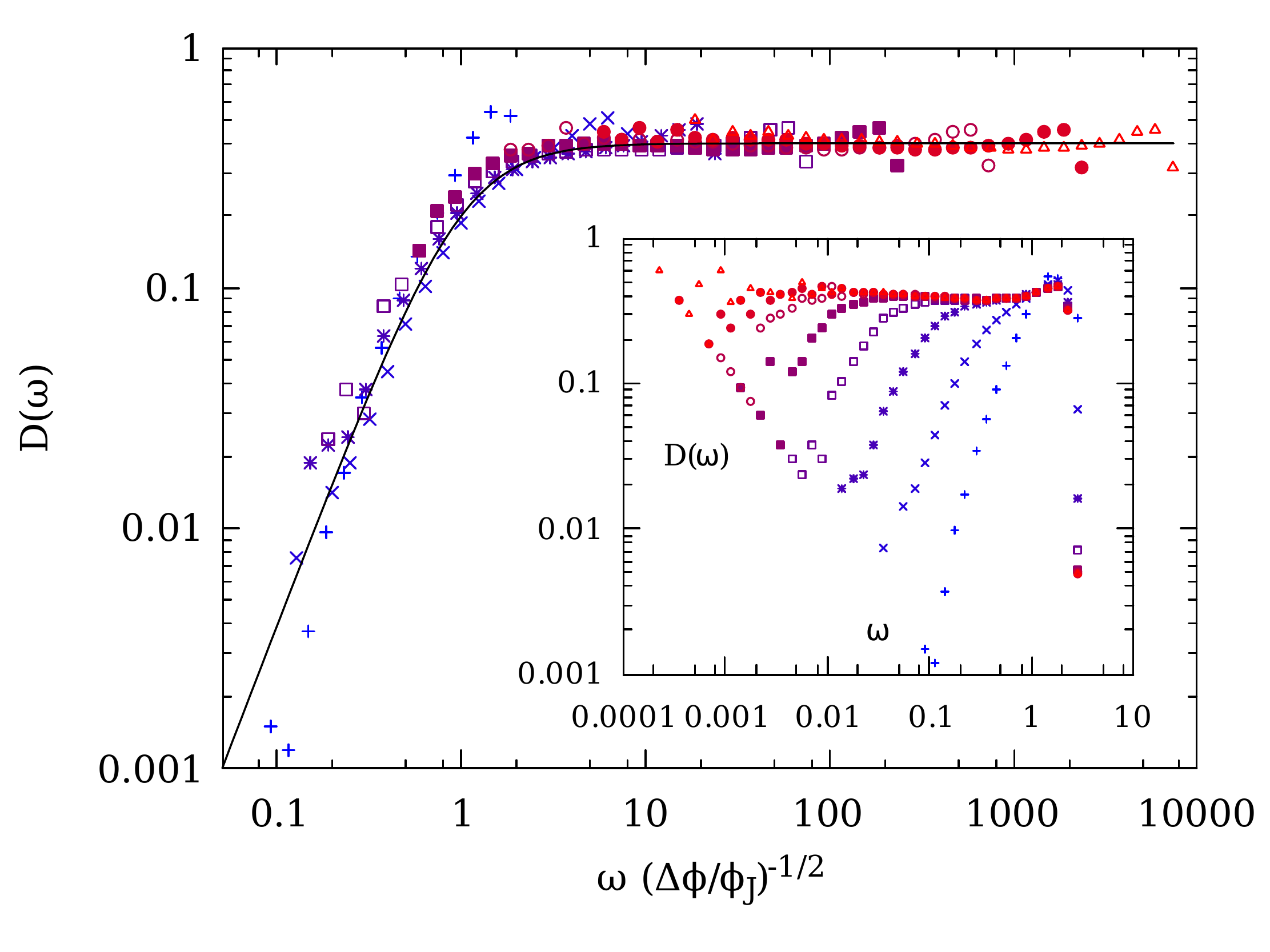} 
\caption{DOS in $d=4$ for $N=8192$ averaged over 50 initial configurations with rescaled $\omega$. The collapsed data is fitted to Eq.~\eqref{eq:fitDOS} (solid line). Inset: DOS for $\Delta\phi/\phi_\mathrm{J}=9.1\times 10^{-8}, 9.2\times 10^{-7}, 9.2\times 10^{-6}, 9.2\times10^{-5}, 9.0\times10^{-4}, 8.7\times10^{-3}, 8.0\times10^{-2}$ and $9.4\times10^{-1}$, from left to right.}
\label{DOS-4d}
\end{figure}	

\paragraph{Universal low-frequency scaling-- } In all dimensions $d$ studied, $D(\omega)$ is found to have the same overall shape. Figure~\ref{DOS-4d} illustrates this universality for $d=4$. At ${\Delta\phi=0}$, the non-trivial part of the DOS, i.e., excluding rattlers, goes to a constant for $\omega\rightarrow0$, as has been widely reported in $d=2$ and 3~\cite{ohern03,silbert05,WSNW05}.
For $\Delta \phi>0$, the DOS peels off from the plateau below a given $\omega_\ast$. It has been argued theoretically~\cite{WSNW05} and observed in $d=3$~\cite{silbert05} that this crossover should scale as $\omega_\ast \propto \sqrt{\Delta\phi}$. Here also, this scaling collapses $D(\omega)$ onto a single master curve at small $\Delta \phi$ (Fig.~\ref{DOS-4d}). 

\begin{figure}
\centering
\includegraphics[width=\linewidth]{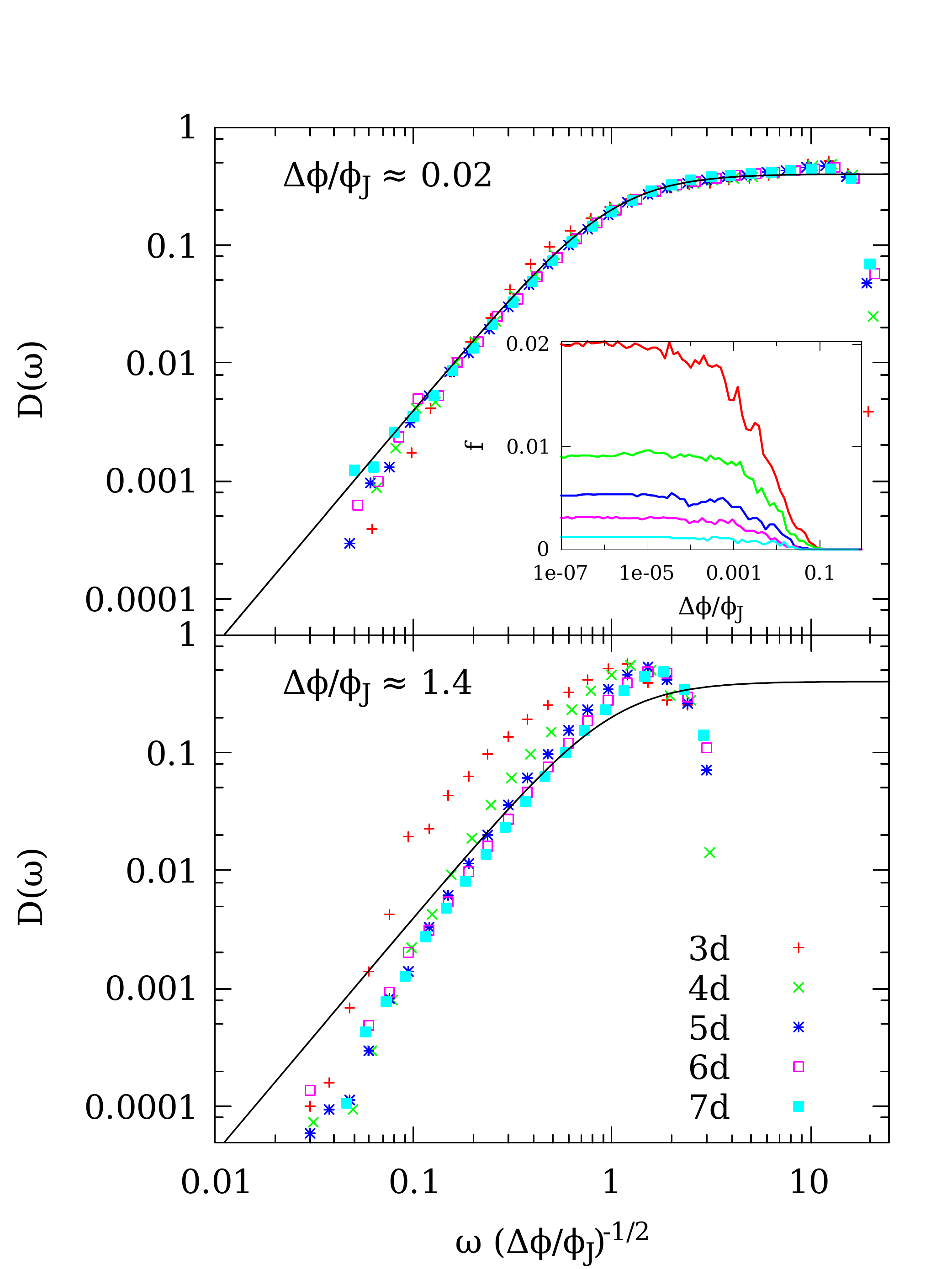}
\caption{DOS for $N=8192$ in $d=3$ to 7, averaged over 30 to 50 configurations, (a) moderately compressed $\Delta \phi/\phi_\mathrm{J}\approx0.2$ and  (b) highly compressed $\Delta\phi/\phi_\mathrm{J}\approx 1.4$ above jamming. The solid line gives Eq.~\eqref{eq:fitDOS} with the fit parameters obtained in Fig.~\ref{DOS-4d}. Inset: The fraction $f$ of rattlers in the system (and hence the fraction of trivial zero modes in the DOS) vanishes around $\Delta\phi\approx 0.1$, and is exponentially suppressed with dimension~\cite{CCPZ12}.}
\label{DOS-AllDims}
\end{figure}

From the rescaled results, we clearly see that below $\omega_\ast$ the DOS scales as $\omega^2$ in all $d$, as in the mean-field descriptions~\cite{degiuli14,FPUZ15}. In order to further scrutinize these results we use the perceptron DOS~\cite{FPUZ15}, 
\begin{equation}
\label{eq:perceptron}
D(\omega ) = \frac{\omega^2(\omega_{\text{max}}^2 - \omega^2)^{1/2}/\pi}{\omega^2 + \omega_\ast^2},
\end{equation}
where $\omega_{\text{max}}$ is the highest frequency.
The model is not translationally invariant and thus its DOS never reaches a Debye-like regime, even as $\omega\ll\omega_0$,
but as long as ${\omega_0\ll\omega \ll \omega_{\text{max}}}$, we can reasonably fit the soft sphere results to a generalized form,
\begin{equation}\label{eq:fitDOS}
D(\omega, \Delta \phi ) \approx \cfrac{A \omega^2}{\omega^2 + B\Delta \phi/\phi_\mathrm{J}},
\end{equation}
with free parameters $A$ and $B$ (Fig.~\ref{DOS-4d}). Remarkably, Eq.~\eqref{eq:fitDOS} captures the collapse of the DOS in all $d$, with roughly the same fitted values for $A \approx 0.4$ and $B \approx 1$ in all cases (Fig.~\ref{DOS-AllDims} top). We thus consistently determine that $\omega_0$ is smaller than
the observable range of frequencies, which is compatible both with a very weak scaling of $\omega_0$ with $\Delta\phi$ and its complete vanishing, as observed in infinite-dimensional models.
	
This scaling universality does not, however, extend to very large $\Delta\phi$. For $\Delta\phi/\phi_\mathrm{J} \gtrsim 1$, it systematically deviates from the master curve on which the lower density results effortlessly collapse (Fig.~\ref{DOS-AllDims}). The low frequency regime then grows faster than $\omega^2$ in all dimensions, but is not Debye-like either. In fact, no clear power-law scaling can be observed. Interestingly, the results nonetheless tend toward a dimensionally-independent form as $d$ increases (Fig. \ref{DOS-AllDims}, bottom), suggesting a certain universality. A possible interpretation is that the various scaling regimes are then mixed, with $\omega_0\sim\omega_\ast\sim\omega_{\mathrm{max}}$, hence the phenomenological form (Eq.~\eqref{eq:fitDOS}) fails. The weak dimensional dependence might result from the integration within the force network of particles that were rattlers at $\phi_\mathrm{J}$, which occurs for $\Delta\phi\approx 0.1$ in low $d$ and decays exponentially quickly with increasing $d$ (Fig. \ref{DOS-AllDims}, inset)~\cite{CCPZ12}. Although we cannot provide a clear resolution of these effects here, we get back to this issue in the conclusion.

\begin{figure}
\centering
\includegraphics[width=\linewidth]{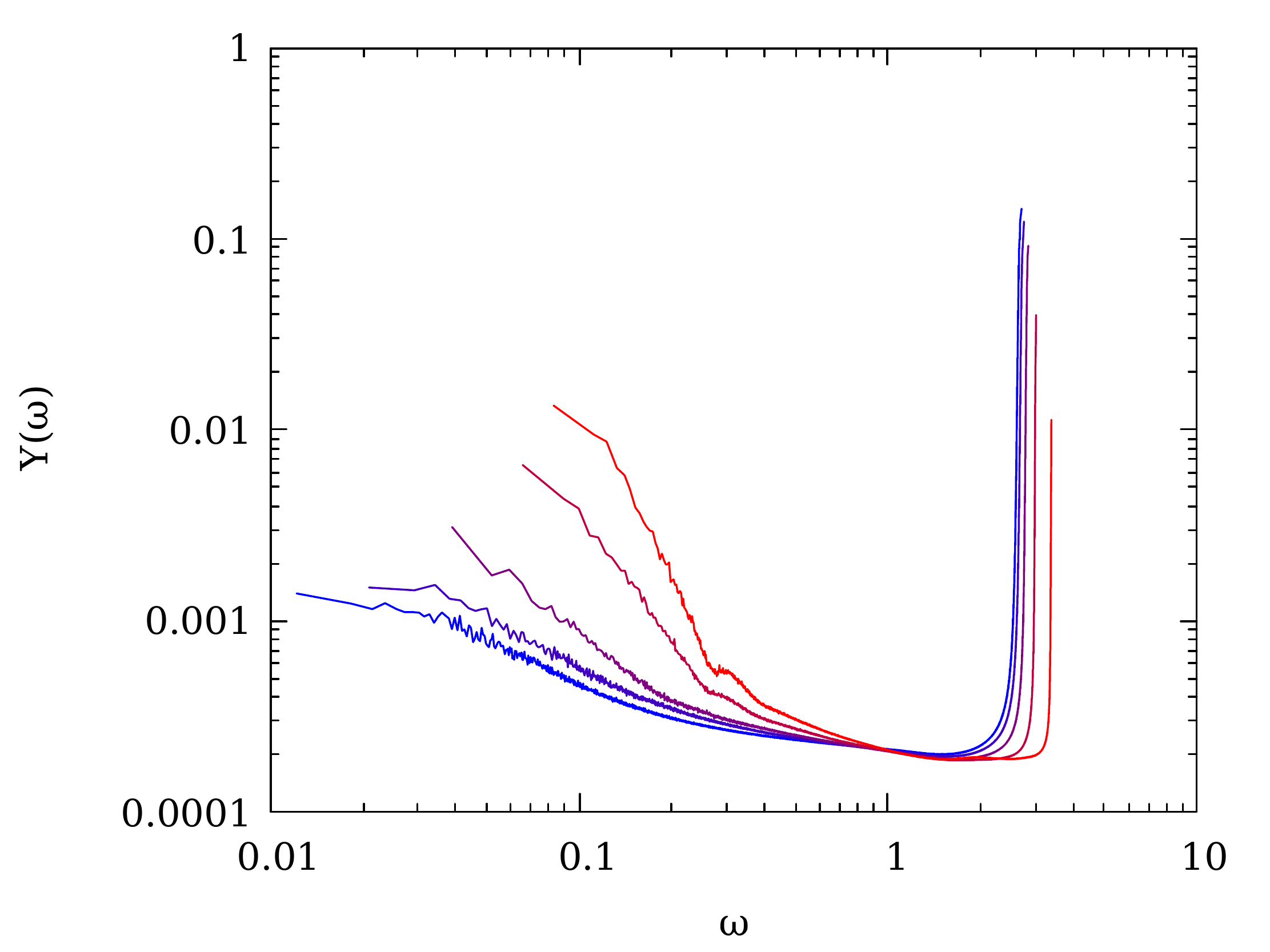}
\caption{Evolution of IPR with frequency in $d=4$ for $N=8192$ averaged over 50 configurations for $\Delta\phi/\phi_\mathrm{J}=1.5\times 10^{-2}, 4.7\times10^{-2}, 1.5\times10^{-1}, 5.1\times10^{-1}$ and 1.6, from left to right.}
\label{IPR-4d}
\end{figure}

\paragraph{Localization--} In light of the remarkable agreement between the mean-field descriptions and soft spheres (for $\Delta\phi/\phi_\mathrm{J}\lesssim 1$), one may wonder if their eigenmode structure is also similar. In $d=2$ and 3, however, soft spheres are known to have low-frequency modes that are quasilocalized~\cite{widmer-cooper08,XVLN10,manning11,manning15}, while modes in the perceptron are known to be perfectly delocalized~\cite{FPUZ15}. Is quasilocalization a feature of soft spheres, or of low-dimensional systems~\cite{parisi14}? In order to identify the source of this difference, we assess the degree of localization of each eigenmode $\{\vec{u}_i(\omega)\}$ by measuring its inverse participation ratio (IPR)~\cite{BMPP15}, 
\begin{equation}
Y(\omega) = \frac{\sum_i^N |\vec{u}_i(\omega)| ^4}{\left[\sum_i |\vec{u}_i(\omega)| ^2\right]^2}.
\end{equation}
By this measure, a mode that is completely localized on a single particle has $Y=1$, while a mode extended over the full system has $Y \sim N^{-1}$. 

\begin{figure}
\centering
\includegraphics[width=\linewidth]{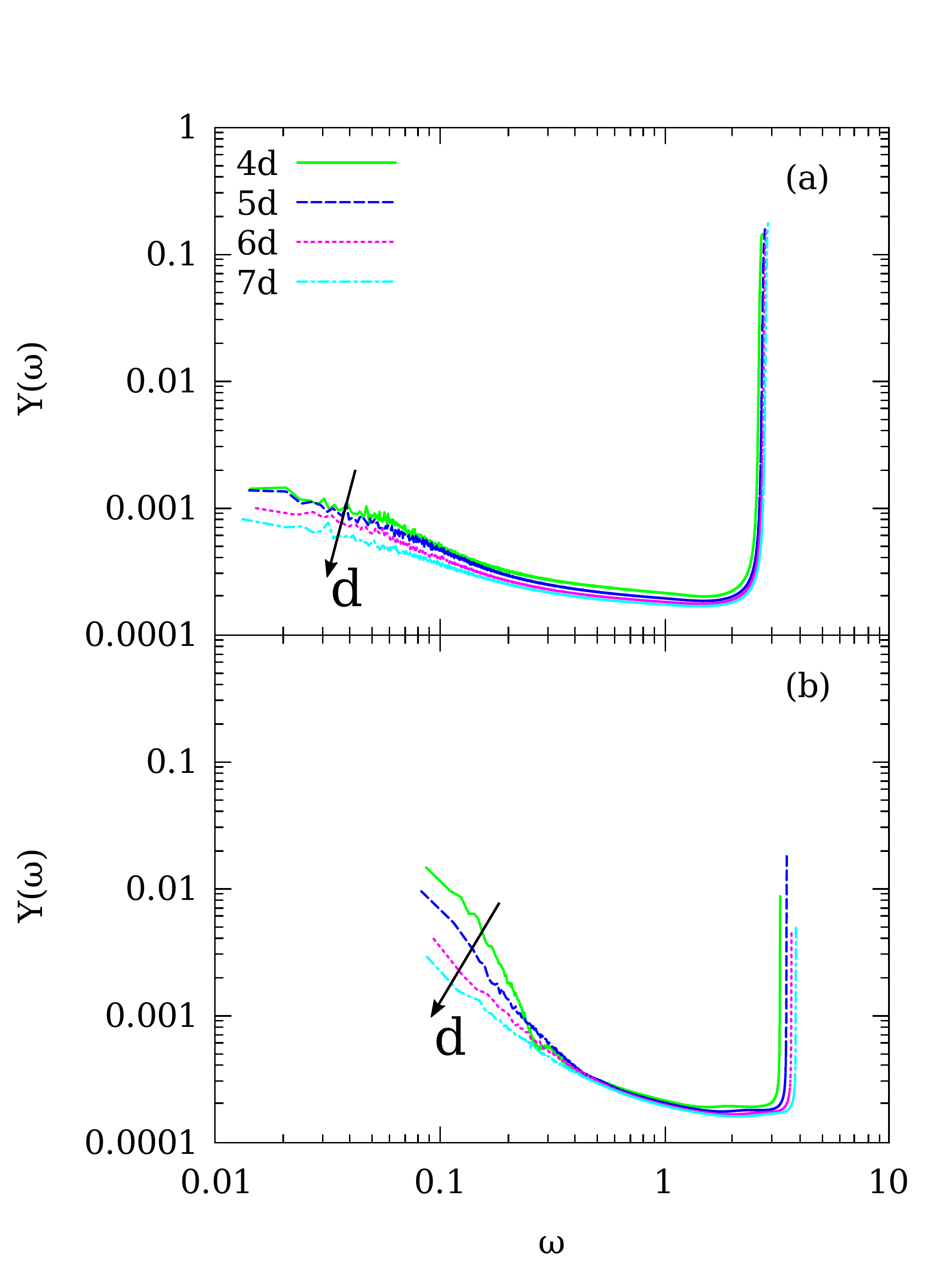}
\caption{IPR for $N=8192$ particles in $d=4$--7, averaged over 30 to 50 configurations, (a) moderately compressed $\Delta \phi/\phi_\mathrm{J}\approx0.2$ and  (b) highly compressed $\Delta\phi/\phi_\mathrm{J}\approx 1.4$ above jamming.}
\label{IPR-AllDims}
\end{figure}

Figure \ref{IPR-4d} shows the evolution of the IPR with frequency in $d=4$ for different $\Delta\phi$. Following Ref.~\cite{XVLN10}, we distinguish three regimes: (i) at low frequency, we find relatively localized modes with intermediate IPR (as in the Heisenberg model~\cite{BMPP15}); 
(ii) at intermediate frequency, we find a band of extended modes with $Y\sim1/N$; and (iii) at high frequency, 
we find Anderson localized modes, as is common in disordered media. In addition to its robust evolution with frequency, $Y(\omega)$ shows an interesting dependence on $\Delta\phi$ and $d$. As $\Delta\phi$ increases, low-frequency modes become increasingly localized (Fig.~\ref{IPR-4d}), while upon approaching jamming these same modes become increasingly delocalized. Delocalization, however, is never complete. In order to observe further IPR decrease, one must consider higher-dimensional systems (Fig.~\ref{IPR-AllDims}). Going from $d=4$ to 7 indeed systematically decreases the degree of localization for all $\Delta\phi$. In other words, the higher the dimension is, the more extended the modes are. Upon reaching $d=\infty$ the spheres are thus expected to behave equivalently to the perceptron. Localized modes are thus clearly a low-dimensional feature. Their precise geometrical origin, however, remains the object of active study~\cite{schoenholz15}.

\paragraph{Conclusion --}
Our analysis reconciles the DOS of amorphous solids from the perceptron and effective medium theory~\cite{FPUZ15,degiuli14}, on the one hand, and simulation results in $d=2$ and 3~\cite{widmer-cooper08,XVLN10}, on the other. The key observations are two-fold: (i) mean-field scaling of the DOS is robustly observed in all dimensions for $\Delta\phi/\phi_\mathrm{J}\lesssim 1$, (ii) delocalization of low-frequency modes increases as $\Delta\phi/\phi_\mathrm{J}\rightarrow 0$ and as $d\rightarrow\infty$. The boson peak is therefore a universal feature of amorphous solids whose origin is purely mean-field in nature, while quasilocalized modes are a low-dimensional effect whose origin is likely related to specific geometrical features~\cite{schoenholz15}. Because some of the structural features of configurations at $\phi_\mathrm{J}$, notably rattlers and bucklers, vanish exponentially as $d$ increases~\cite{CCPZ12,CCPZ15}, we tentatively conclude that quasilocalized modes also cannot be obtained perturbatively from mean-field, infinite-dimensional descriptions. We also conclude that the existence of a boson peak is independent of quasilocalization proper, in contrast to Ref.~\cite{manning15}.

Another interesting feature of the DOS is observed for $\Delta\phi/\phi_\mathrm{J}\gtrsim 1$. Even though the results remain largely independent of dimension, they are quite distinct from the mean-field scaling form that easily describes the $\Delta \phi/\phi_\mathrm{J}<1$ regime. Whether this effect is due to a breakdown of some of the assumptions made in the comparison or to the presence of a phase transition at $T=0$ as has recently been proposed~\cite{BU15} remains, however, an open question.

\begin{acknowledgments}
The simulations were run on the University of Oregon's ACISS supercomputer (156 NVDIA M2070).
We wish to thank P.~Urbani for stimulating discussions. PC acknowledges support from the National Science Foundation 
Grant No.~NSF DMR-1055586 and the Sloan Foundation. E.I.C. thanks the NSF for support under CAREER Award No. DMR- 1255370.  The use of the ACISS supercomputer is supported under a Major Research Instrumentation Grant, Office of Cyber Infrastructure, No. OCI-0960354.
\end{acknowledgments}

\bibliographystyle{apsrev4-1}
\bibliography{jamming,HS} 
\end{document}